\documentclass[dvips]{acta}
\usepackage{supertabular,lscape,epsfig}
\usepackage{amssymb}
\usepackage{amsmath}
\SetPages{1}{8}     % ***
\SetVol{67}{2017}

\begin{document}

\begin{Titlepage}

\Title { DW UMa and the Irradiation Modulated Mass Transfer Model \\ for Superhumps }

\Author {J.~~S m a k}
{N. Copernicus Astronomical Center, Polish Academy of Sciences,\\
Bartycka 18, 00-716 Warsaw, Poland\\
e-mail: jis@camk.edu.pl }

\Received{  }

\end{Titlepage}

\Abstract { The light curves of the permanent superhumper DW UMa are analyzed 
in order to determine the amplitudes of its superhumps, $A_{SH}$, 
and the amplitudes of the periodic light variations with the beat period -- 
the {\it irradiation amplitudes}, $A_{irr}$. 
The resulting values of $A_{SH}$ and $A_{irr}$, together with other values 
from the literature, turn out to be correlated thereby confirming the irradiation 
modulated mass transfer model for superhumps. 
}
{\it accretion, accretion disks, binaries: cataclysmic variables, 
stars: individual: DW UMa }

%Sec.1
\section { Introduction } 

Superhumps are periodic light variations with periods slightly longer than 
the orbital period observed in dwarf novae during their superoubursts and  
also in some cataclysmic variables with stationary accretion -- 
the so-called {\it permanent superhumpers} (Warner 2003, and references therein). 
They have been discovered during the 1972 superoutburst of the dwarf nova VW Hyi  
(Vogt 1974, Warner 1975), 
but first observed (although unrecognized as such at that time) 
ten years earlier in AM CVn (Smak 1967). 
One of the permanent superhumpers is the nova-like, eclipsing system DW UMa. 

There are two, competing models for superhumps: the tidal-resonance (TR) model and 
the irradiation modulated mass transfer (IMMT) model. 
The TR model (reviewed in Section 2) fails to explain many important facts. 
The IMMT model (Section 3) appears to be very promising and the aim of the present 
paper is to test one of its crucial ingredients. The light curves of DW UMa are 
analyzed in Section 4 and the test is presented in Section 5.

%Sec.2
\section { The Tidal-Resonance Model for Superhumps }   

The tidal-resonance (TR) model, first proposed by Whitehurst (1988) 
and Hirose and Osaki (1990), explains superhumps as being due 
to tidal effects in the outer parts of accretion disks, leading -- {\it via} the 3:1 
resonance -- to the formation of an eccentric outer ring undergoing apsidal motion 
and periodic dissipation of the kinetic energy of disk's elements. 
This model and, in particular, the results of numerous 2D and 3D SPH 
simulations (cf. Pearson 2006, Smith et al. 2007 and references therein) 
reproduce (although not without problems) the observed superhump periods; 
this suggests that the basic "clock" which 
defines the superhump periods may indeed be related to the apsidal motion. 

On the other hand, however, the TR model fails to explain  
many other important facts (cf. Smak 2010). In particular: 

{\parskip=0truept {
(1) The 3:1 resonance, which is the crucial ingredient of this model, can occur only 
in systems with mass ratios $q<q_{crit}=0.25$.  
This condition is not fulfilled by longer period CV's, including the dwarf nova U Gem 
and the growing number of permanent superhumpers 
(one of them is DW UMa with $q=0.39\pm 0.12$; Araujo-Betancor et al. 2003). 

(2) The numerical 2D and 3D SPH simulations produce "superhumps" with amplitudes 
which -- compared to the observed amplitudes  -- are about 10 times too small 
(Smak 2009a). 
}}

In spite of that the TR model continues to be commonly accepted... 

%Sec.3
\section { The Irradiation Modulated Mass Transfer Model for Superhumps } 

From the analysis of eclipses observed in Z Cha (Smak 2007,2009b, based on light 
curves collected by Warner and O'Donoghue 1988) it was found that 
(1) the standard hot spot can be detected only during eclipses which occur away 
from superhump maximum; 
(2) during eclipses which occur closer to the superhump maximum it is replaced 
by the "hot line" or "hot strip" resulting from stream overflow; and 
(3) the superhump light source is located along the stream trajectory. 

Further clue came from the analysis of the superoutburst light curves of several 
dwarf novae: it was found (Smak 2009c) that in the case of deeply eclipsing systems 
($i>82^\circ$) the observed brightness of the disk (excluding eclipses and superhumps) 
is modulated with phase of the beat period, related to the orbital and superhump 
periods by 

%Eq.1
\beq
{1\over{P_b}}~=~{1\over{P_{orb}}}~-~{1\over{P_{SH}}} .
\eeq

\noindent
The updated version of the original amplitude {\it vs.} inclination 
diagram is shown here in Fig.1. 

%***Fig.1
\begin{figure}[htb]
\epsfysize=10.0cm 
\hspace{1.0cm}
\epsfbox{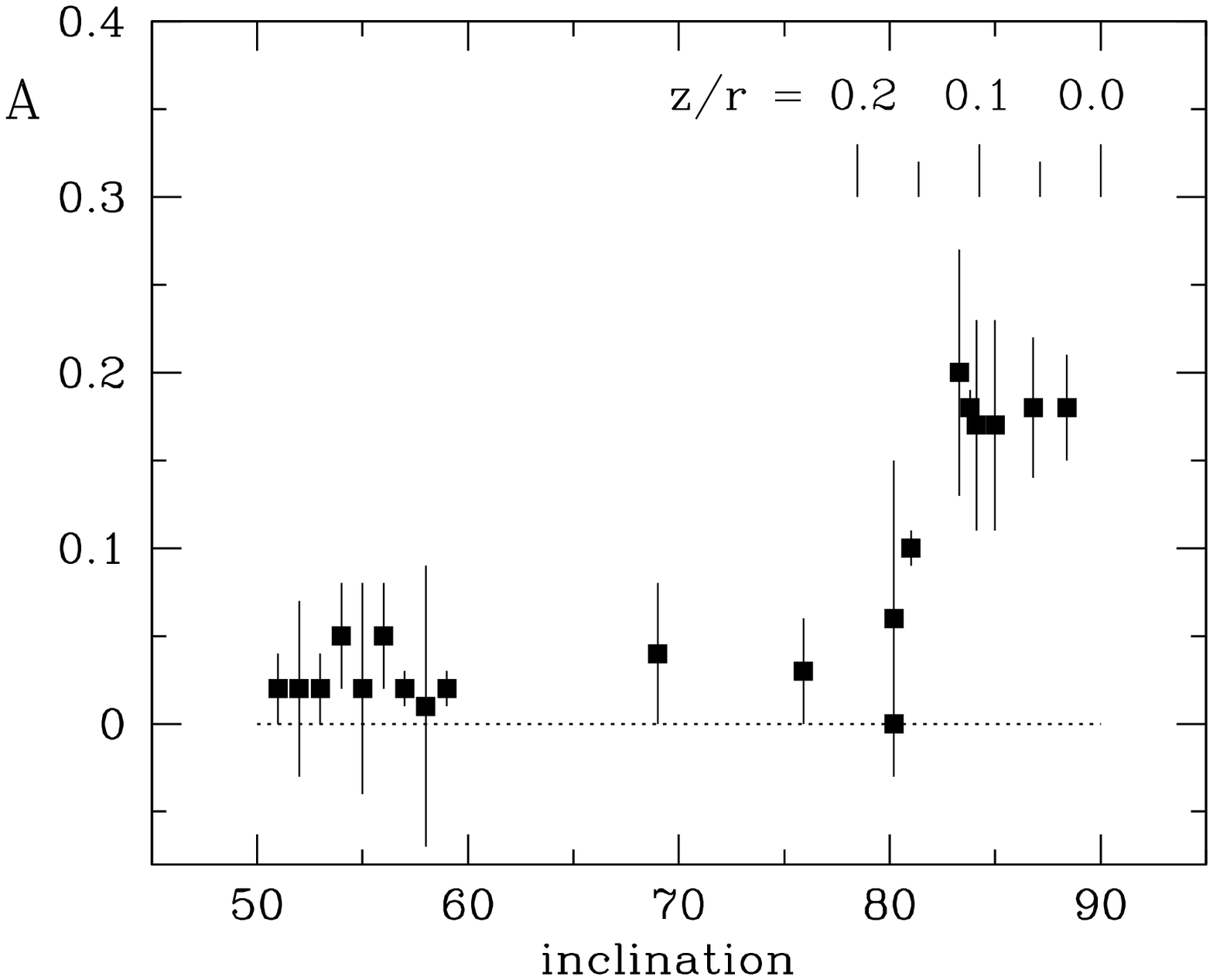} 
\vskip -20truemm
\FigCap { The modulation amplitude as a function of inclination. 
Data are taken from Smak(2009c, Table 2, and 2013, Table 1). 
Non-eclipsing systems with unknown inclinations are plotted between $i=50^\circ$ 
and $60^\circ$. Marked above are values of $z/r=\cos i$.    
}
\end{figure}

The presence of such modulations was interpreted as being due to a non-axisym- metric 
structure of the outer parts of the disk, involving azimuthal dependence of their 
geometrical thickness, rotating with the beat (apsidal motion) period. 

Regardless of this interpretation it was obvious that the periodically variable 
{\it irradiation of the observer} with the beat period implies the periodically 
variable {\it irradiation of the secondary} with the superhump period. 
Taking this into account we will refer to the light variations with $P_b$ -- 
as the {\it irradiation light curve} and to its amplitude -- as the 
{\it irradiation amplitude}. 

The evidence described above led to the irradiation modulated mass tranfer (IMMT) 
model for superhumps (Smak 2009c). It consists of the following essential points: 

{\parskip=0truept {
(1) The irradiation of the secondary component is modulated with $P_{SH}$.  

(2) Periodically variable irradiation of the secondary results in periodic 
variations of the mass transfer rate.   

(3) The periodically variable dissipation of the kinetic energy of the stream  
is observed in the form of superhumps. 

(4) Around superhump maximum the stream overflows the surface of the disk 
and -- unlike in the case of the "standard" hot spot -- its energy is dissipated 
along its trajectory above (and below) the disk. 
}}
\parskip=12truept 

The IMMT model is based on purely observational evidence. 
The only exception is point (2), which remains hypothetical. 
The problem of how the rate of the mass outflow from the secondary can be controlled 
by its irradiation is quite complex due primarily to the fact that the vicinity of 
the inner Lagrangian point L1 remains in the shadow cast by the disk. 
Results of preliminary model calculations (Osaki and Meyer 2003, Smak 2004, Viallet 
and Hameury 2007) were inconlusive and even controversial. The most recent results 
by Cambier (2015) are much more promissing, but it is clear that further work is 
needed before this problem can be solved. 
In this situation the only alternative is to look for direct observational evidence 
supporting this point. This will be done in the next two Sections.

%Sec.4
\section { DW UMa } 

% 4.1
\subsection {DW UMa as Permanent Superhumper} 

DW UMa is an eclipsing, nova-like, permanent superhumper. 
In addition to the positive and negative superhumps (Patterson et al. 2002, 
Boyd et al. 2017 and references therein) it shows spectroscopic peculiarities 
characteristic for the SW Sex stars (Thorstensen et al. 1991, Dhillon et al. 2013, 
and references therein) and occasionally displays the so-called low states 
(Honeycutt et al. 1993, Stanishev et al. 2004). 

The most recent paper by Boyd et al. (2017) presented results of a large photometric 
program covering several seasons. Two of those results are of particular interest 
in the context of the present paper:  
(1) The superhump and irradiation amplitudes are variable. 
(2) The results from 2014 and 2015, when both amplitudes were determined, suggest 
that they are correlated. 
Such a correlation would provide the crucial test for the IMMT model 
(see Section 3). It is therefore of the utmost importance to confirm its existence. 

% 4.2
\subsection {The Data}   

Results by Boyd et al. (2017) show that the superhump period and the superhump 
and irradiation amplitudes are variable. Taking this into account it was decided 
to analyze light curves covering relatively short intervals of time. 
Three such data sets could be recovered from the literature. 

{\bf 1997}. The data used in our analysis come from visual (V) light curves 
observed by B{\'i}r{\'o} (2000) during five consecutive nights of February 16 
-- February 20, 1997. The data points were read from his Fig.3 at equal intervals 
$\Delta \phi_{orb}=0.05$, excluding points with $\phi_{orb}<0.2$ and 
$\phi_{orb}>0.8$, and converted to magnitudes. 
The periodogram of those data showed two significant signals: at the orbital 
frequency (which was removed prior to further analysis) and at 
$f_{SH}=6.76\pm 0.14$ c/d, corresponding to $P_{SH}=0.148\pm 0.003$ d. 

{\bf 2002}. Visual (V) light curves, observed by Stanishev et al. (2004) 
on five nights between January 11 and February 17, 2002, were used with 
data points being read from their Fig.1 at equal intervals $\Delta t=0.005$d, 
excluding points with $\phi_{orb}<0.2$ and $\phi_{orb}>0.8$. 

{\bf 2003}. "Unfiltered" light curves, observed by Stanishev et al. (2004) 
on five nights between March 6 and March 22, 2003, were used 
with data points being obtained in the same way as for 2002. 

% 4.3
\subsection { The superhump and irradiation amplitudes } 

The superhump and irradiation amplitudes, $A_{SH}$ and $A_{irr}$, are 
determined simultaneously -- via the least squares solution -- by fitting 

% ***
% \vskip -10truemm

%Eq.2
\beq
m~=~<m>~+~{{dm}\over{dt}}\Delta t
   ~-~A_{SH}\cos(\phi_{sh}-\phi_{SH}^{max})
   ~-~A_{irr}\cos(\phi_{irr}-\phi_{irr}^{max})~,
\eeq

% ***
% \vskip -5truemm

\noindent 
to individual points. The results are listed in Table 1. 
Listed in that table are also the superhump and irradiation amplitudes taken from 
Boyd and Gaensicke (2009) and from Boyd et al. (2017, Table 3).

% ***Table 1
\begin{table}[h!]
{\parskip=0truept
\baselineskip=0pt {
\medskip
\centerline{Table 1}
\medskip
\centerline{ Superhump Amplitudes and Irradiation Amplitudes }
\medskip
$$\offinterlineskip \tabskip=0pt
\vbox {\halign {\strut
\vrule width 0.5truemm #&	%1
\quad#\quad&                    %2
\vrule#&			%3
\quad\hfil#\hfil\quad&          %4
\vrule#&			%5
\quad#\quad&                    %6
\vrule#&			%7
\quad#\quad&                    %8
\vrule#&			%9
\quad#\hfil\quad&               %10
\vrule width 0.5 truemm # \cr	%11
\noalign {\hrule height 0.5truemm}
&&&&&&&&&&\cr
& Year &&JD 2400000+ &&\hfil $A_{SH}$\hfil &&\hfil $A_{irr}$\hfil && &\cr
&&&&&&&&&&\cr
\noalign {\hrule height 0.5truemm}
&&&&&&&&&&\cr
&1997&&50496-500&&$0.030\pm0.011$&&$0.025\pm0.012$&&(1)&\cr
&2002&&52286-323&&$0.058\pm0.013$&&$0.079\pm0.019$&&(2)&\cr
&2003&&52703-721&&$0.070\pm0.013$&&$0.086\pm0.012$&&(2)&\cr
&2008&&54570-600&&$0.055\pm0.010$&&$0.060\pm0.010$&&(3)&\cr
&2014&&56728-768&&$0.049\pm0.002$&&$0.043\pm0.002$&&(4)&\cr
&2015&&57020-110&&$0.063\pm0.003$&&$0.062\pm0.003$&&(4)&\cr
&&&&&&&&&&\cr
\noalign {\hrule height 0.5truemm}
}}$$
Notes to Table 1: (1) This paper; data from B{\'i}r{\'o} (2000). 
(2) This paper; data from Stanishev et al. (2004). 
(3) Boyd and Gaensicke (2009). (4) Boyd et al. (2017).  
}}
\end{table}

The superhump amplitudes determined for 2002 and 2003 require some comments. 
In the case of 2002 our value $A_{SH}=0.058\pm0.013$, based on light curves between 
JD 2452286-323, is larger than $A_{SH}=0.048$ obtained by Stanishev et al. (2004) 
from light curves between JD 2452286-389, and $A_{SH}=0.045\pm0.004$ obtained by 
Boyd et al. (2017) from light curves between JD 2452311-332. 
On the other hand, however, the amplitude determined from the three nights between 
JD 2452373-389 (Stanishev et al. 2004, Fig.1) is lower: $A_{SH}=0.035\pm0.013$  
(regretfully, the coverage in $\phi_{irr}$ was insufficient for simultenaous 
determination of $A_{irr}$). 
This shows that during the 2002 season the superhump amplitude was simply decreasing. 
In the case of 2003 our value $A_{SH}=0.070\pm0.013$, based on light curves between 
JD 2452703-721, is larger than $A_{SH}=0.040$ obtained by Stanishev et al. (2004) 
from light curves between JD 2452703-793. Our determination, based on the same interval, 
gave similar value: $A_{SH}=0.036\pm0.016$.

% 4.4
\subsection { The Light Curves }  

The superhump and irradiation light curves are then obtained as

% ***
% \vskip -5truemm

%Eq.3
\beq
\Delta m_{SH}~=~m~-~[~<m>~+~{{dm}\over{dt}}\Delta t 
     ~-~A_{irr}\cos(\phi_{irr}-\phi_{irr}^{max})~]~, 
\eeq

% ***
% \vskip -5truemm

\noindent
with a similar expression for $\Delta m_{irr}$. They are shown in Fig.2.

%***Fig.2
\begin{figure}[htb]
\epsfysize=9.0cm 
\hspace{2.0cm}
\epsfbox{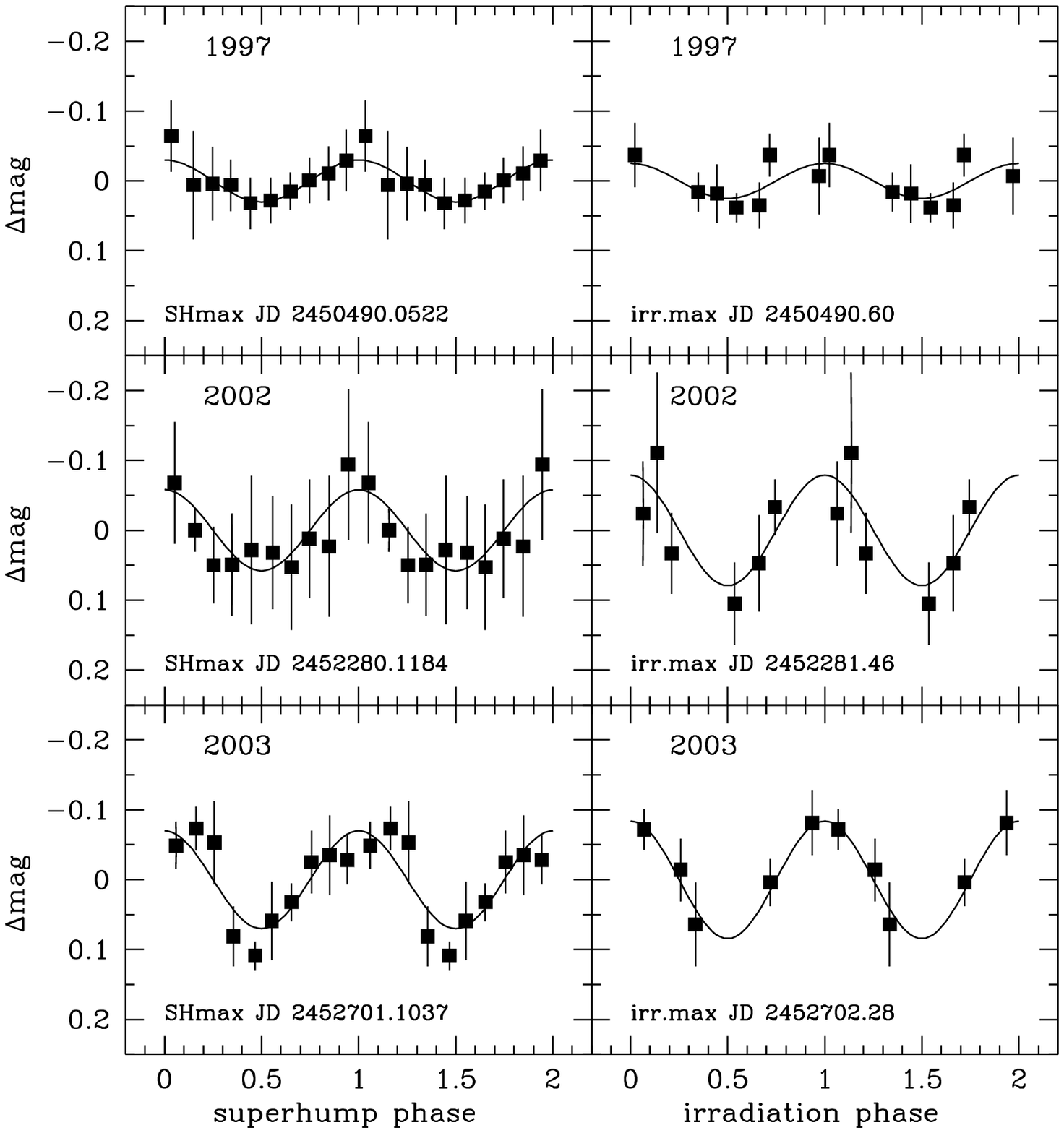} 
\vskip -2truemm
\FigCap { Superhump (left) and irradiation (right) light curves of DW UMa 
in 1997, 2002, and 2003. Normal points are shown with error bars representing 
the scatter of individual points. 
}
\end{figure}

% Sec.5
\section { The Crucial Test for the IMMT Model }  

The superhump amplitudes and the irradiation amplitudes from Table 1 
are compared in Fig.3. They are correlated. The formal fit to the points gives 

%Eq.4 
\beq
A_{SH}~=~(0.34\pm0.07)~A_{~irr}^{(0.62\pm0.07)}~.	                 
\eeq

%***Fig.3
\begin{figure}[htb]
\epsfysize=10.0cm 
\hspace{1.0cm}
\epsfbox{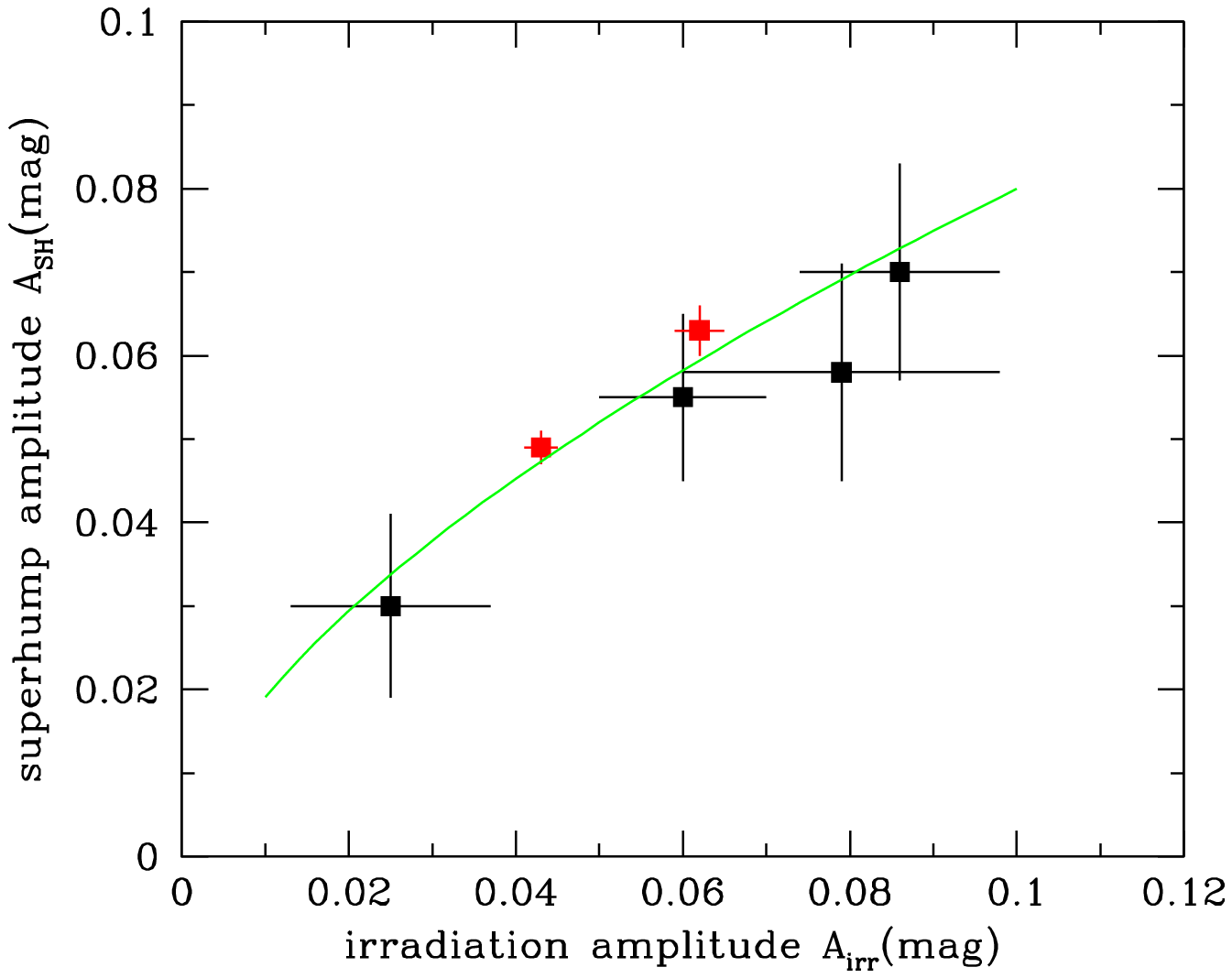} 
\vskip -35truemm
\FigCap { The dependence of superhump amplitudes on irradiation amplitudes. 
Data points and their errors are taken from Table 1. 
Sympols in red represent the two original points from Boyd et al. (2017). 
Green line represents Eq.4. 
}
\end{figure}

To conclude: the superhump amplitude does indeed depend on the irradiation 
amplitude thereby confirming 
the last element of the IMMT model which still required such a confirmation. 
It can only by hoped that further such tests involving either DW UMa or other 
permanent superhumpers will strengthten this conclusion.

\begin {references} 

\refitem {Araujo-Betancor, S. et al.} {2003} {\ApJ} {583} {437} 

\refitem {B{\'i}r{\'o}, I.B.} {2000} {\AA} {364} {573}

\refitem {Boyd, D., Gaensicke, B.} {2009} 
         {\it Proceedings for the 28th Annual Conference of the Society 
         for Astronomical Sciences} {28} {127}

\refitem {Boyd, D.R.S. et al.} {2017} {\MNRAS} {466} {3417}

\refitem {Cambier, H.} {2015} {\MNRAS} {452} {3620}

\refitem {Dhillon, V.S., Smith, D.A., Marsh, T.R.} {2013} {\MNRAS} {428} {3559}

\refitem {Hirose, M., Osaki, Y.} {1990} {\em Publ.Astr.Soc.Japan} {42} {135}

\refitem {Honeycutt, R.K., Livio, M., Robertson, J.W.} {1993} {\PASP} {105} {922} 

\refitem {Osaki, Y., Meyer, F.} {2003} {\AA} {401} {325} 

\refitem {Patterson, J. et al.} {2002} {\PASP} {114} {1364} 

\refitem {Pearson, K.J.} {2006} {\MNRAS} {371} {235}

\refitem {Smak, J.} {1967} {\Acta} {17} {255}

\refitem {Smak, J.} {2004} {\Acta} {54} {181}

\refitem {Smak, J.} {2007} {\Acta} {57} {87}

\refitem {Smak, J.} {2009a} {\Acta} {59} {103}

\refitem {Smak, J.} {2009b} {\Acta} {59} {109}

\refitem {Smak, J.} {2009c} {\Acta} {59} {121}  

\refitem {Smak, J.} {2010} {\Acta} {60} {357}

\refitem {Smak, J.} {2013} {\Acta} {63} {369}

\refitem {Smith, A.J., Haswell, C.A., Murray, J.R., Truss, M.R., Foulkes, S.B.} 
         {2007} {\MNRAS} {378} {785}

\refitem {Stanishev, V., Kraicheva, Z., Boffin, H.M.J., Genkov, V., 
          Papadaki, C., Carpano, S.} {2004} {\AA} {416} {1057}

\refitem {Thorstensen, J.R., Ringwald, F.A., Wade, R.A., Schmidt, G.D., 
          Norsworthy, J.E.} {1991} {\AJ} {102} {272} 

\refitem {Viallet, M., Hameury, J.-M.} {2007} {\AA} {475} {597} 

\refitem {Vogt, N.} {1974} {\AA} {36} {369} 

\refitem {Warner, B.} {1975} {\MNRAS} {170} {219}

\refitem {Warner, B.} {2003} {\it Cataclysmic Variable Stars, 2nd edition,  
         Cambridge University Press} {~} {~}. 

\refitem {Warner, B., O'Donoghue, D.} {1988} {\MNRAS} {233} {705} 

\refitem {Whitehurst, R.} {1988} {\MNRAS} {232} {35} 

\end {references}

\end{document}